\documentclass[11pt]{article}
\usepackage{graphicx}
\usepackage{longtable,lscape}
\usepackage{amsfonts}
\usepackage{float}
\usepackage{url}

\newcommand{\captionfonts}{\footnotesize}
\makeatletter  % Allow the use of @ in command names
\long\def\@makecaption#1#2{%
  \vskip\abovecaptionskip
  \sbox\@tempboxa{{\captionfonts #1: #2}}%
  \ifdim \wd\@tempboxa >\hsize
    {\captionfonts #1: #2\par}
  \else
    \hbox to\hsize{\hfil\box\@tempboxa\hfil}%
  \fi
  \vskip\belowcaptionskip}
\makeatother 
\begin{document}
\title{Quantum structure in cognition: Why and \\ how concepts are entangled}
\author{Diederik Aerts and Sandro Sozzo \vspace{0.5 cm} \\ 
        \normalsize\itshape
        Center Leo Apostel for Interdisciplinary Studies \\
        \normalsize\itshape
        Brussels Free University \\ 
        \normalsize\itshape
         Krijgskundestraat 33, 1160 Brussels, Belgium \\
        \normalsize
        E-Mails: \url{diraerts@vub.ac.be,ssozzo@vub.ac.be}
        }
\date{}
\maketitle 
\begin{abstract}
\noindent One of us has recently elaborated a theory for modelling concepts that uses the state context property (SCoP) formalism, i.e. a generalization of the quantum formalism. This formalism incorporates context into the mathematical structure used to represent a concept, and thereby models how context influences the typicality of a single exemplar and the applicability of a single property of a concept%. The notion of `state of a concept' is introduced to account for this contextual influence
, which provides a solution of the {\it Pet-Fish problem} and other difficulties occurring in concept theory. Then, a quantum model has been worked out which
reproduces the membership weights of several exemplars of concepts and their combinations. We show in this paper that a further relevant effect appears in a natural way whenever two or more concepts combine, namely, {\it entanglement}. The presence of entanglement is explicitly revealed by considering a specific example with two concepts, constructing some Bell's inequalities for this example, testing them in a real experiment with test subjects, and finally proving that Bell's inequalities are violated in this case. We show that the intrinsic and unavoidable character of entanglement can be explained in terms of the weights of the exemplars of the combined concept with respect to the weights of the exemplars of the component concepts.
\end{abstract}


\section{Introduction\label{intro}}
Understanding the mechanism of how concepts combine to form sentences and texts so that it is possible to communicate meaning among human minds is one of the major challenges in the psychological studies on human thought. None of the existing theories on concepts explains however `how concepts combine'. This {\it combination problem} was manifestly revealed by Hampton's experiments \cite{hampton88a,hampton88b} which measured the deviation from classical set theoretic membership weights of exemplars with respect to pairs of concepts and their conjunction or disjunction. Hampton's investigation was motivated by the so-called {\it Guppy effect} in concept conjunction found by Osherson and Smith \cite{oshersonsmith81}. These authors considered the concepts {\it Pet} and {\it Fish} and their conjunction {\it Pet-Fish}, and observed that, while an exemplar such as {\it Guppy} was a very typical example of {\it Pet-Fish}, it was neither a very typical example of {\it Pet} nor of {\it Fish}. Hence, the typicality of a specific exemplar with respect to the conjunction of concepts shows an unexpected behavior from the point of view of classical set and probability theory. As a result of this work, the problem is often referred to as the {\it Pet-Fish problem} and the effect is usually called the {\it Guppy effect}. Hampton identified a Guppy-like effect for the membership weights of exemplars with respect to pairs of concepts and their conjunction \cite{hampton88a}, and equally so for the membership weights of exemplars with respect to pairs of concepts and their disjunction \cite{hampton88b}. Several experiments have since been performed (see, e.g., \cite{hampton01}) and many approaches have been propounded to solve the Pet-Fish problem (see, e.g., fuzzy set based theories \cite{zadeh01,zadeh02,oshersonsmith02}) and to provide a satisfactory mathematical model of concept combinations (see, e.g., explanation based theories \cite{komatsu01,fodor01,rips03}). But none of the currently existing concept theories provide a satisfactory description or explanation of such effects  \cite{hampton01,fodor01,rips03}.  

Inspired by a formalism providing an operational foundation of quantum mechanics \cite{aerts1986,aerts1992,aerts1994,aerts1999b}, one of the authors has elaborated, together with some co-workers, a {\it State Context Property} ({\it SCoP}) formalism to model and represent concepts \cite{gaboraaerts02,aertsgabora2005a,aertsgabora2005b,aertsczachordhooghe2006}. In the SCoP formalism each concept is associated with well defined sets of states, contexts and properties. Concepts change continuously under the influence of a context and this change is described by a change of the state of the concept. For each exemplar of a concept, the typicality varies with respect to the context that influences it. Analogously, for each property, the applicability varies with respect to the context. This implies the presence of both a {\it contextual typicality} and an {\it applicability effect}. The Pet-Fish problem is solved in the SCoP formalism because in different combinations the concepts are in different states. In particular, in the combination {\it Pet-Fish} the concept {\it Pet} is in a state under the context {\it The Pet is a Fish}. The state of {\it Pet} under the context {\it The Pet is a Fish} has different typicalities, which explains the Guppy effect. On the basis of the SCoP formalism, a mathematical model using the formalism of quantum mechanics in Hilbert space has been worked out which allows one to reproduce the experimental results obtained by Hampton on conjunctions and disjunctions of concepts. This formulation identifies the presence of typically quantum effects in the mechanism of combination of concepts, e.g., contextual influence, superposition, interference, emergence, etc. \cite{aerts2009a,aerts2009c,aertsaertsgabora2009,aertsdhooghe2009,aertsdhooghehaven2010,aerts2007a,aerts2007b}.

In this paper we show that another relevant effect which is usually considered as characteristic of quantum mechanical entities, that is, {\it entanglement}, is present whenever two or more concepts combine. The presence of entanglement is explicitly revealed by considering two concepts, i.e. {\it Animal} and {\it Acts}, and their combination {\it The Animal Acts}, together with some exemplars {\it Horse}, {\it Bear}, {\it Tiger}, {\it Cat} (for {\it Animal}) and {\it Growls}, {\it Whinnies}, {\it Snorts}, {\it Meows} (for {\it Acts}), and constructing some Bell's inequalities in the version derived by Clauser, Horne, Shimony and Holt \cite{chsh69} (Sec. \ref{theoretical}). We then test these Bell's inequalities in a real experiment with 81 test subjects and analyze the obtained data (Sec. \ref{experiment}). The experiment shows a significant violation of Bell's inequalities, hence it proves the entanglement between the concepts {\it Animal} and {\it Acts} when they form the sentence {\it The Animal Acts} (by the term {\it entanglement} we actually mean the presence of nonclassical correlations violating Bell's inequalities, without reference to any mathematical representation in Hilbert spaces). Moreover, we compare the obtained data with the results that would have been obtained if context and meaning had not influenced the subjects' minds. In the latter case, indeed, Bell's inequalities are not violated, hence their violation in our experiment shows that meaning and context play a basic role in the combination of concepts. We finally provide an explanation of the origins and ubiquity of entanglement in combined concepts in terms of weights of the exemplars of the combined concept with respect to the weights of the exemplars of the component concepts (Sec. \ref{explanation}).

We conclude this section by observing that the potentially fundamental role played by entanglement in word association was pointed out by Nelson and McEvoy and Bruza et al. in \cite{nelsonmcvoy07,bruzaetal08}. In \cite{bruzaetal09} it is shown that if one assumes that words can become entangled in the human mental lexicon, then one can provide a unified framework in which two seemingly competing approaches for modeling the activation level of words in human memory, namely, the {\it Spreading Activation} and the {\it Spooky-activation-at-a-distance}, can be recovered. 

\section{Detecting entanglement between concepts\label{theoretical}}
We illustrate in this section how entanglement appears in a natural way whenever two or more concepts combine. To this aim, we analyze here an example with two concepts and a combination along the lines put forward in \cite{aertsgabora2005a,aertsgabora2005b,aertsczachordhooghe2006}.

We regard the sentence {\it The Animal Acts} as a conceptual entity, hence as a combination of the concepts {\it Animal} and {\it Acts}. Then, we show the presence of entanglement between these two concepts by testing Bell's inequality with respect to them. We consider two couples of exemplars or states of the concept {\it Animal}, namely {\it Horse}, {\it Bear} and {\it Tiger}, {\it Cat}, and also two couples of exemplars or states of the concept {\it Acts}, namely {\it Growls}, {\it Whinnies} and {\it Snorts}, {\it Meows} -- for our experiment we specifically consider forms of actions, hence exemplars of {\it Acts}, which consists of possible animal sounds, hence exemplars of {\it Making A Sound}. Our first experiment $A$ consists in test subjects choosing between the two exemplars {\it Horse} and {\it Bear} to answer the question `is a good example of' the concept {\it Animal}, and we put $E(A)=+1$ if {\it Horse} is chosen, hence the state of {\it Animal} changes to {\it Horse}, and $E(A)=-1$ if {\it Bear} is chosen, hence the state of {\it Animal} changes to {\it Bear}, introducing in this way the function $E$ which measures the `expectation value' for the test outcomes concerned. Our second experiment $A'$ consists in test subjects choosing between the two exemplars {\it Tiger} and {\it Cat} to answer the question `is a good example of' the concept {\it Animal}, and we consistently put $E(A')=+1$ if {\it Tiger} is chosen and $E(A')=-1$ if {\it Cat} is chosen to introduce a measure of the expectation value. The third experiment $B$ consists in test subjects choosing between the two exemplars {\it Growls} and {\it Whinnies} to answer the question `is a good example of' the concept {\it Acts}, with $E(B)=+1$ if {\it Growls} is chosen and $E(B)=-1$ if {\it Whinnies} is chosen. The fourth experiment $B'$ consists in test subjects choosing between the exemplars {\it Snorts} and {\it Meows} to answer the question `is a good example of' the concept {\it Acts}, with $E(B')=+1$ if {\it Snorts} is chosen and $E(B')=-1$ if {\it Meows} is chosen.

Let us now consider coincidence experiments in combinations $AB$, $A'B$, $AB'$ and $A'B'$ for the conceptual combination {\it The Animal Acts}. Concretely, this means that, for example, test subjects taking part in the experiment $AB$, to answer the question `is a good example of', will choose between the four possibilities (1) {\it The Horse Growls}, (2) {\it The Bear Whinnies} -- and if one of these is chosen we put $E(AB)=+1$ -- and (3) {\it The Horse Whinnies}, (4) {\it The Bear Growls} -- and if one of these is chosen we put $E(AB)=-1$. For the coincidence experiment, $A'B$ subjects, to answer the question `is a good example of', will choose between (1) {\it The Tiger Growls}, (2) {\it The Cat Whinnies} -- and in case one of these is chosen we put $E(A'B)=+1$ -- and (3) {\it The Tiger Whinnies}, (4) {\it The Cat Growls} -- and in case one of these is chosen we put $E(A'B)=-1$. For the coincidence experiment, $AB'$ subjects, to answer the question `is a good example of', choose between (1) {\it The Horse Snorts}, (2) {\it The Bear Meows} -- and in case one of these is chosen we put $E(AB')=+1$ -- and (3) {\it The Horse Meows}, (4) {\it The Bear Snorts} -- and in case one of these is chosen we put $E(AB')=-1$. And finally, for the coincidence experiment, $A'B'$ subjects, to answer the question `is a good example of', will choose between (1) {\it The Tiger Snorts}, (2) {\it The Cat Meows} -- and in case one of these is chosen we put $E(A'B')=+1$ -- and (3) {\it The Tiger Meows}, (4) {\it The Cat Snorts} -- and in case one of these is chosen we put $E(A'B')=-1$. We can now evaluate the expectation values $E(A', B')$, $E(A', B)$, $E(A, B')$ and $E(A,B)$ associated with the coincidence experiments $A'B'$, $A'B$, $AB'$ and $AB$, respectively, and substitute them into the Clauser-Horne-Shimony-Holt variant of Bell's inequality \cite{chsh69} 
\begin{equation} \label{chsh}
-2 \le E(A',B')+E(A',B)+E(A,B')-E(A,B) \le 2.
\end{equation}
From the well-known analysis of Bell's inequality follows that in case the experimental expectation values violate Eq. (\ref{chsh}), a local and classical probabilistic description is not possible, and entanglement exists between the given concepts. Thus, by the sentence Animal {\it is entangled with} Acts we mean the experimental fact that these two concepts exhibit nonclassical correlations, without referring to any mathematical representation in Hilbert spaces. The connections with entangled states in tensor product Hilbert spaces will be outlined in Sec. \ref{explanation}.

We note that the maximum violation of the Bell's inequality in Eq. (\ref{chsh}) occurs when the quantity $ E(A',B')+E(A',B)+E(A,B')-E(A,B)$ is equal to +4, that is, when the outcome for each one of the members of this expression is +1, +1, +1 and -1, respectively. Let us make an intuitive analysis of the situation such that we can see why Bell's inequality will most probably be violated for our experiment. In the coincidence experiment $AB$, both {\it The Horse Whinnies} and {\it The Bear Growls} will yield rather high scores, with the two remaining possibilities {\it The Horse Growls} and {\it The Bear Whinnies} being chosen little. This means that we will get $E(A,B)$ close to -1. On the other hand, in the coincidence experiment $A'B$ one of the four choices will be prominent, namely {\it The Tiger Growls}, while the three other possibilities, {\it The Cat Whinnies}, {\it The Tiger Whinnies}, and {\it The Cat Growls}, will be much less present amongst the choices made by the test subjects. This means that we have $E(A',B)$ close to +1. In the two remaining coincidence experiments, we equally have that only one of the choices is prominent. For $A,B'$, this is {\it The Horse Snorts}, with the other three {\it The Bear Meows}, {\it The Horse Meows} and {\it The Bear Snorts} being much less present. For $A'B'$, the prominent choice is {\it The Cat Meows}, while the other three {\it The Tiger Snorts}, {\it The Tiger Meows} and {\it The Cat Snorts} are much less present. This means that we have $E(A,B')$ is close to +1 and $E(A',B')$ is close to +1. Coming to the expectation values, we hence can expect that Eq. (\ref{chsh}) be violated, and that case (ii) occurs such that the existence of entanglement between the considered concepts would be proven. 

One of us has recently shown \cite{aerts2010} that Eq. (\ref{chsh}) is violated in the concept combination {\it The Animal Acts} by using the World Wide Web as a conceptual domain. In the next section we will show that a violation occurs also when the data are collected from a real experiment with test subjects following the standard procedure of psychology experiments in concept research.

\section{Description of the experiment\label{experiment}}
The entanglement mentioned in the foregoing section was tested in an experiment where 81 participating subjects were presented with a questionnaire to be filled out accompanied by the following text:

%\vspace{.1cm}
{\it This study has to do with what we have in mind when we use words that refer to categories, and more specifically `how we think about examples of categories'. Let us illustrate what we mean. Consider the category `fruit'. Then `orange' and `strawberry' are two examples of this category, but also `fig' and `olive' are examples of the same category. In each test of the questionnaire you will be asked to pick one of the examples of a set of given examples for a specific category. And we would like you to pick that example that you find `a good example' of the category. In case there are more than one example which you find a good example, pick then the one you find the best of all the good examples. In case there are two examples which you both find equally good, and hence hesitate which ones to take, just take then the one you slightly prefer, however slight the preference might be. It is mandatory that you always `pick one and only one example', hence in case of doubt, anyhow pick one and only one example. This is necessary for the experiment to succeed. So, one of the tests could be that the category `fruit' is given, and you are asked to pick one of the examples `orange', `strawberry', `fig' or `olive' as a good example, and in case of doubt the best of the ones you doubt about, and in case you cannot decide, pick one anyhow. Let all aspects of yourself play a role in the choice you make, ratio, but also imagination, feeling, emotion, and whatever.}

%\vspace{.1cm}

Let us now examine the obtained results.

For the coincidence experiment $AB$, $4$ subjects chose the example {\it The Horse Growls} as a good example of the combination {\it The Animal Acts}, $5$ subjects chose {\it The Bear Whinnies}, $51$ subjects chose {\it The Horse Whinnies}, and $21$ subjects chose {\it The Bear Growls}. This means that on a totality of $81$ test subjects we get fractions of $4$, $5$, $51$ and $21$ for the different combinations considered. This allows us to calculate the probability for one of the combinations to be chosen. We have $P(A_1,B_1)=4/81=0.0494$ for {\it The Horse Growls}, $P(A_2,B_2)=21/81=0.2593$ for {\it The Bear Whinnies}, $P(A_1,B_2)=51/81=0.6296$ for {\it The Horse Whinnies} and $P(A_2,B_1)=5/81=0.0617$ for {\it The Bear Growls}. Knowing these probabilities, we can again calculate the expectation value for this coincidence experiment by means of the equation $E(A,B)=P(A_1,B_1)+P(A_2,B_2)-P(A_2,B_1)-P(A_1,B_2)=-0.7778$. We calculate the expectation values $E(A',B)$, $E(A,B')$ and $E(A',B')$ in an analogous way. For the coincidence experiment $A'B$, $63$ subjects chose the example {\it The Tiger Growls} as a good example of the combination {\it The Animal Acts}, $4$ subjects chose {\it The Cat Whinnies}, $7$ subjects chose {\it The Tiger Whinnies}, and  $7$ subjects chose {\it The Cat Growls}. This gives $P(A'_1,B_1)=0.7778$, $P(A'_2,B_2)=0.0494$, $P(A'_1, B_2)=0.0864$ and $P(A'_2,B_1)=0.0864$, hence $E(A',B)=0.6543$. For the coincidence experiment $AB'$, $48$ subjects chose the example {\it The Horse Snorts} as a good example of the combination {\it The Animal Acts}, $7$ subjects chose {\it The Bear Meows}, $2$ subjects chose {\it The Horse Meows}, and $24$ subjects chose {\it The Bear Snorts}. This gives $P(A_1,B'_1)=0.5926$, $P(A_2,B'_2)=0.0864$, $P(A_1, B'_2)=0.0247$ and $P(A_2,B'_1)=0.2963$, hence $E(A,B')=0.3580$. For the coincidence experiment $A'B'$, $12$ subjects chose the example {\it The Tiger Snorts} as a good example of the combination {\it The Animal Acts}, $54$ subjects chose  {\it The Cat Meows}, $7$ subjects chose {\it The Tiger Meows}, and $8$ subjects chose {\it The Cat Snorts}. This gives $P(A'_1,B'_1)=0.1481$, $P(A'_2,B'_2)=0.6667$, $P(A'_1, B'_2)=0.0864$ and $P(A'_2,B'_1)=0.0988$, hence $E(A',B')=0.6296$. For the expression appearing in the Clauser-Horne-Shimony-Holt variant of Bell's inequalities, we get
\begin{equation} \label{bell}
E(A',B')+E(A',B)+E(A,B')-E(A,B)=2.4197
\end{equation}
which is manifestly greater than 2, hence it violates Bell's inequalities and reveals entanglement between the concept {\it Animal} and the concept {\it Acts}. 

The above violation of Bell's inequalities constitutes our main result in this paper and we will exhaustively comment on it in the next section. But we first want to consider Bell's inequalities under different perspectives. 

Suppose that there are two separated sources of knowledge, e.g., two test subjects, and consider the coincidence experiment $AB$. Let $P(A_1)$ be the probability that the first subject choose the exemplar {\it Horse} as a good example of the concept {\it Animal}, let $P(B_1)$ be the probability that the second subject choose the exemplar {\it Growls} as a good example of the concept {\it Acts}, and let us estimate the probability that the example {\it The Horse Growls} be a good example of the conceptual combination {\it The Animal Acts} as the product $P(A_1)P(B_1)$, that is, as the joint probability $P_{\mathit{prod}}(A_1,B_1)$ that the first subject choose {\it Horse} and the second subject choose {\it Growls}. By referring to the experimental data that have been collected we have $P(A_1)=43/81=0.5309$, $P(B_1)=39/81=0.4815$, $P_{\mathit{prod}}(A_1,B_1)=P(A_1)P(B_1)=0.2556$. Analogously, we can calculate the probability that {\it The Bear  Whinnies} be a good example of {\it The Animal Acts} as the product of the probability $P(A_2)$ that the first subject choose {\it Bear} as a good example of {\it Animal} times the probability $P(B_2)$ that the second subject choose {\it Whinnies} as a good example of {\it Acts}. We find from empirical data $P(A_2)=38/81=0.4691$, $P(B_2)=42/81=0.5185$, hence $P_{\mathit{prod}}(A_2,B_2)=P(A_2)P(B_2)=0.2433$. By proceeding in an analogous way we calculate the probability that {\it The Horse Whinnies} be a good example of {\it The Animal Acts} as the product of the probability $P(A_1)$ that the first subject choose {\it Horse} as a good example of {\it Animal} times the probability $P(B_2)$ that the second subject choose {\it Whinnies} as a good example of {\it Acts}. We find $P_{\mathit{prod}}(A_1,B_2)=P(A_1)P(B_2)=0.2753$. Furthermore, if we calculate the probability that {\it The Bear Growls} be a good example of {\it The Animal Acts} as the product of the probability $P(A_2)$ that the first subject choose {\it Bear} as a good example of {\it Animal} times the probability $P(B_1)$ that the second subject choose {\it Growls} as a good example of {\it Acts}, we find $P_{\mathit{prod}}(A_2,B_1)=P(A_2)P(B_1)=0.2259$. The expectation value is $E_{\mathit{prod}}(A,B)= P_{\mathit{prod}}(A_1,B_1)+P_{\mathit{prod}}(A_2,B_2)-P_{\mathit{prod}}(A_2,B_1)-P_{\mathit{prod}}(A_1,B_2)=-0.0022$. Let us now consider the coincidence experiment $A'B$. The probability that the first subject choose the example {\it Tiger} as a good example of {\it Animal} is $P(A'_{1})=59/81=0.7284$, while the probability that the first subject choose {\it Cat} as a good example of {\it Animal} is $P(A'_{2})=22/81=0.2716$. If we calculate the probability that {\it The Tiger Growls} be a good example of {\it The Animal Acts} as the product of the probability $P(A'_{1})$ that the first subject choose {\it Tiger} as a good example of {\it Animal} times the probability $P(B_1)$ that the second subject choose {\it Growls} as a good example of {\it Acts}, we find $P_{\mathit{prod}}(A'_1,B_1)=P(A'_{1})P(B_1)=0.3507$. Analogously, we find $P_{\mathit{prod}}(A'_2,B_2)=P(A'_{2})P(B_2)=0.1408$, $P_{\mathit{prod}}(A'_1,B_2)=P(A'_{1})P(B_2)=0.3777$, $P_{\mathit{prod}}(A'_2,B_1)=P(A'_{2})P(B_1)=0.1308$. The expectation value is $E_{\mathit{prod}}(A',B)=P_{\mathit{prod}}(A'_1,B_1)+P_{\mathit{prod}}(A'_2,B_2)-P_{\mathit{prod}}(A'_2,B_1)-P_{\mathit{prod}}(A'_1,B_2)=-0.0169$. Let us come to the coincidence experiment $AB'$. The probability that the second subject choose the example {\it Snorts} as a good example of {\it Acts} is $P(B'_{1})=26/81=0.3210$, while the probability that the second subject choose {\it Meows} as a good example of {\it Acts} is $P(B'_{2})=55/81=0.6790$. If we calculate the probability that {\it The Horse Snorts} be a good example of {\it The Animal Acts} as the product of the probability $P(A_{1})$ that the first subject choose {\it Horse} as a good example of {\it Animal} times the probability $P(B'_1)$ that the second subject choose {\it Snorts} as a good example of {\it Acts}, we find $P_{\mathit{prod}}(A_1,B'_1)=P(A_{1})P(B'_1)=0.1704$. Analogously, we find $P_{\mathit{prod}}(A_1,B'_1)=P(A_{2})P(B'_2)=0.3185$, $P_{\mathit{prod}}(A_1,B'_1)=P(A_{1})P(B'_2)=0.3605$, $P_{\mathit{prod}}(A_1,B'_1)=P(A_{2})P(B'_1)=0.1506$. The expectation value is $E_{\mathit{prod}}(A,B')=P_{\mathit{prod}}(A_1,B'_1)+P_{\mathit{prod}}(A_2,B'_2)-P_{\mathit{prod}}(A_2,B'_1)-P_{\mathit{prod}}(A_1,B'_2)=-0.0221$. Finally, let us consider the coincidence experiment $A'B'$. If we calculate the probability that {\it The Tiger Snorts} be a good example of {\it The Animal Acts} as the product of the probability $P(A'_{1})$ that the first subject choose {\it Tiger} as a good example of {\it Animal} times the probability $P(B'_1)$ that the second subject choose {\it Snorts} as a good example of {\it Acts}, we find $P_{\mathit{prod}}(A'_1,B'_1)=P(A'_{1})P(B'_1)=0.2338$. Analogously, we find $P_{\mathit{prod}}(A'_2,B'_2)=P(A'_{2})P(B'_2)=0.1844$, $P_{\mathit{prod}}(A'_1,B'_2)=P(A'_{1})P(B'_2)=0.4946$, $P_{\mathit{prod}}(A'_2,B'_1)=P(A'_{2})P(B'_1)=0.0871$. The expectation value is $E_{\mathit{prod}}(A',B')=P_{\mathit{prod}}(A'_1,B'_1)+P_{\mathit{prod}}(A'_2,B'_2)-P_{\mathit{prod}}(A'_2,B'_1)-P_{\mathit{prod}}(A'_1,B'_2)=-0.1635$. The `product' expectation values $E_{\mathit{prod}}(A,B)$, $E_{\mathit{prod}}(A',B)$, $E_{\mathit{prod}}(A,B')$ and $E_{\mathit{prod}}(A',B')$ can then be put into the Bell inequality, which gives
\begin{equation}
E_{\mathit{prod}}(A',B')+E_{\mathit{prod}}(A',B)+E_{\mathit{prod}}(A,B')-E_{\mathit{prod}}(A,B)=-0.2003.
\end{equation}
This result is very different from the earlier obtained expression, and also does not violate Bell's inequalities. The reason for this is that in the case of `separated sources of knowledge', the non-violation of Bell's inequalities is structural \cite{aerts2010}. %Let us show this by first proving the following lemma: 
This statement can be proved as follows.

\noindent
{\it Lemma. If $x$, $x'$, $y$ and $y'$ are real numbers such that $-1\le x, x', y , y\le +1$ and $S=xy + xy' + x'y - x'y'$ then $-2\le S \le +2$.} 

\noindent
{\it Proof.} Since $S$ is linear in all four variables $x$, $x'$, $y$, $y'$, it must take on its maximum and minimum values at the corners of the domain of this quadruple of variables, that is, where each of $x$, $x'$, $y$, $y'$ is +1 or -1. Hence at these corners $S$ can only be an integer between -4 and +4. But $S$ can be rewritten as $(x + x')(y + y') - 2x'y'$, and the two quantities in parentheses can only be 0, 2, or -2, while the last term can only be -2 or +2, so that S cannot equal -3, +3, -4, or +4 at the corners. 

Since in the situation considered we have $P_{prod}(A_i,B_j)=P(A_i)P(B_j)$, $P_{prod}(A'_i,B_j)=P(A'_i)P(B_j)$, $P_{prod}(A_i,B'_j)=P(A_i)P(B'_j)$ and $P_{prod}(A'_i,B'_j)=P(A'_i)P(B'_j)$, we get $E(A,B)=E(A)E(B)$, $E(A',B)=E(A')E(B)$, $E(A,B')=E(A)E(B')$ and $E(A',B')=E(A')E(B')$, and hence from the lemma it follows that $-2\le E(A'B')+E(A'B)+E(AB')-E(AB) \le +2$, which proves the Clauser-Horne-Shimony-Holt variant of Bell's inequalities to be valid.

The foregoing considerations show that one of the elements in the violation of Bell's inequalities is the non-product nature of the probabilities $P(A_i,B_j)$, $P(A'_i,B_j)$, $P(A_i,B'_j)$ and $P(A'_i,B'_j)$, e.g., $P(A_i,B_j) \ne P(A_i)P(B_j)$. If we understand why these coincidence probabilities are not of the product nature we can get an insight into one of the elements of the violation of Bell's inequalities for the situations that we have considered. 
Indeed, consider for example the probability $P(A_1,B_1)$ and let us analyze why it is different from $P(A_1)P(B_1)$. We have that $P(A_1,B_1)$ is the probability, empirically estimated, that a given test subject choose the sentence {\it The Horse Growls} as a good example of the concept {\it The Animal Acts}, and then we find $P(A_1,B_1)=0.0494$. On the contrary, $P(A_1)P(B_1)$ is the probability that, of two given test subjects, the first choose {\it Horse} as a good example of {\it Animal} and the other choose independently {\it Growls} as a good example of {\it Acts}, and then we find $P(A_1)P(B_1)=0.2556$. These values are very different. Indeed, the probability to find the sentence part {\it The Horse Growls} is little, for any meaning this sentence may have will not be easily ascertained, since it is most unusual for horses to growl. If however two `separated' or `independent' subjects are chosen at random, the probability that {\it Horse} be chosen by one of them, and {\it Growls} be chosen by the other, is substantial. The fundamental reason for this difference is that in the second case the choices are `separated' or `independent' or, rather, `not connected by meaning'.

The results above show that `meaning' plays a fundamental role in determining the experimental weights of the examples of concept combinations. But, there are stronger arguments to maintain that context and meaning are crucial in human thought, hence a combination of concepts is not like a `bag of words', as implied by the mathematical structure of existing semantic theories, e.g., LSA.

Let us calculate the data that would have been obtained if the minds of the test subjects had not been influenced by context and meaning. Consider the coincidence experiment $AB$ and suppose that a given subject chooses the exemplar {\it Horse} as a good example of the concept {\it Animal} and {\it Growls} as a good example of the concept {\it Acts}. Should context and meaning not play any role, then the subject would choose with certainty the example {\it The Horse Growls} as a good example of the combination {\it The Animal Acts}. We can thus evaluate the probability $P_{\mathit{class}}(A_1,B_1)$ that a given subject choose {\it Horse} in the experiment $A$ and {\it Growls} in the experiment $B$. It is given by $P_{\mathit{class}}(A_1,B_1)=19/81=0.2346$, where $19$ is the number of subjects who chose {\it Horse} in the experiment $A$ and {\it Growls} in the experiment $B$. This probability can be used as an estimation of the probability that a given subject choose {\it The Horse Growls} as a good example of {\it The Animal Acts}. We can repeat the same reasoning for the other possible results in the coincidence experiment $AB$, thus getting  $P_{\mathit{class}}(A_2,B_2)=0.2222$, $P_{\mathit{class}}(A_1,B_2)=0.2963$ and $P_{\mathit{class}}(A_2,B_1)=0.2469$. Hence the expectation value is $E_{\mathit{class}}(A, B)=P_{\mathit{class}}(A_1,B_1)+P_{\mathit{class}}(A_2,B_2)-P_{\mathit{class}}(A_1,B_2)-P_{\mathit{class}}(A_2,B_1)=-0.0864$ in this case. Analogously, we get $E_{\mathit{class}}(A',B)=0.1235$, $E_{\mathit{class}}(A, B')=-0.0123$ and $E_{\mathit{class}}(A',B')=-0.1111$ for the expectation values of the other coincidence experiments. The `classical' expectation values $E_{\mathit{class}}(A,B)$, $E_{\mathit{class}}(A',B)$, $E_{\mathit{class}}(A,B')$ and $E_{\mathit{class}}(A',B')$ can then be inserted into the Bell inequality, which gives
\begin{equation}
E_{\mathit{class}}(A',B')+E_{\mathit{class}}(A',B)+E_{\mathit{class}}(A,B')-E_{\mathit{class}}(A,B)=0.0864.
\end{equation}
As we can see, the obtained value does not violate Bell's inequalities. As a consequence, the violation of Bell's inequalities in the experiment that we have considered can be interpreted as proving that meaning and context are fundamental for the mechanism of construction of sentences.

To conclude this section we observe that we also performed a statistical analysis of the empirical data using the `t-test for paired two samples for means' to estimate the probability that the shifts from Bell's inequalities be due to chance. We compared the data collected in the real experiment with the data collected in the `classical' experiment, where no influence of context and meaning is present. For the 16 pairs to compare the p-values came out as follows: 0.000392657, 0.003921785, 2.50665E-06, 0.820174295, 3.8846E-08, 0.011513803, 4.78134E-05, 0.741136115, 2.35428E-08, 0.000152291, 1.3612E-08, 0.006518053, 0.073431676, 7.38957E-12, 3.8846E-08, 0.56693215. This makes it possible to conclude convincingly that the deviation effects are not caused by random fluctuations.

\section{Explanation of entanglement in concepts\label{explanation}}
A fundamental consequence of the experimental results obtained in Sec. \ref{experiment} is that any formalism aiming at representing concepts should incorporate the possibility of having entangled concepts from the very beginning. In order to understand in depth the mechanism of entanglement between concepts together with the causes of its ubiquity, we put forward an analysis of the situation in this section with the aim of grasping the core element of entanglement for concept combinations. 

Consider the concept {\it Animal}. This concept is an `abstraction' of possible concrete exemplars of {\it Animal}, e.g., {\it Horse}, {\it Bear}, {\it Tiger}, {\it Cat}, etc. When we ask a subject to estimate whether a given example, say, {\it Horse} is a `good example' of the concept {\it Animal} this operation corresponds to `wandering into the realm of abstraction and concretization'. The concept {\it Animal} is then connected with the exemplars of {\it Animal} by weights, expressing frequencies of appearance and/or typicalities of the different exemplars. Analogously, the concept {\it Acts} is an abstraction of possible concrete exemplars of {\it Acts}, and also connected to these different exemplars by weights, expressing frequencies of appearance and/or typicalities. Let us now consider the concept {\it The Animal Acts} which is the combination of {\it Animal} and {\it Acts}. This is also an abstraction of possible exemplars. In the situation that we considered for the experiment the concrete exemplars are {\it The Horse Growls}, {\it The Tiger Meows}, etc.. But, the weight of, say, {\it The Horse Growls} is not the product of the weight of {\it Horse} in {\it Animal} times the weight of {\it Growls} in {\it Acts} in this case, otherwise Bell's inequalities would have been satisfied. It follows that the essential element being at the origin of entanglement is that `when concepts combine they do this inside the realm of where they exist as abstractions'. With other words, the combination {\it The Animal Acts}, is a combination of two abstractions {\it Animal} and {\it Acts}, but it does not connect with the concrete elements, i.e. the exemplars of {\it Animal} and {\it Acts}. No, it connects with its own set of exemplars, such a {\it The Horse Whinnies} or {\it The Bear Growls}, etc., which are in themselves combinations of exemplars of the original concepts, but even this is not necessary, also completely new exemplars can be considered for the combination. This is a very different way of combining than for example the way in which two classical physical object combine. Hence, entanglement is a direct and deep consequence of this special way of combining, for each combination choosing its own set of new exemplars, `with new weight specifically linked to the individual exemplars', and not connecting to the product set of the old exemplars and corresponding weights. That concepts have this special way of combining in common with quantum entities might not be a coincidence, a hypothesis investigated in \cite{aerts2009a}.

A consequence of the above analysis is that entanglement in concepts does not strictly depend on the linearity of the tensor product Hilbert space that can be used to model the entity {\it The Animal Acts} -- we remind that the {\it Tsirelson inequalities} \cite{tsirelson80} hold in the specific case that we have considered, therefore a purely quantum model can be worked out in this case. %However, the presence of a quantum model can be useful to represent mathematically how concepts entangle. More concretely, 
Moreover, the type of model in Hilbert space that we would expect is the following. Let us denote the states of the concepts {\it Animal} and {\it Acts} by the unit vectors $|p_{\mathit{Animal}}\rangle$ and $|p_{\mathit{Acts}}\rangle$, respectively. Since {\it Animal} and {\it Acts} are both abstractions of, say, {\it Horse} and {\it Bear} and of {\it Growls} and {\it Whinnies}, respectively, we have
\begin{equation}
|p_{\mathit{Animal}}\rangle = a_1 |p_{H}\rangle + a_2 |p_{B}\rangle, \quad |p_{\mathit{Acts}}\rangle = b_1 |p_{G}\rangle+b_2 |p_{W}\rangle
\end{equation}
where $|a_1|^2$ and $|a_2|^2$, and $|b_1|^2$ and $|b_2|^2$, respectively, are the weights that both concretizations carry, and the unit vectors $|p_{H}\rangle$, $|p_{B}\rangle$, $|p_{G}\rangle$ and $|p_{W}\rangle$ represent the states $p_{\mathit{Horse}}$, $p_{\mathit{Bear}}$, $p_{\mathit{Growls}}$ and $p_{\mathit{Whinnies}}$, respectively. The ground state $p_{\mathit{The\, Animal\, Acts}}$ of the combination {\it The Animal Acts}, being an abstraction of `all combinations of the concrete cases', is then represented by the unit vector
\begin{equation}
|p_{\mathit{The\, Animal\, Acts}}\rangle=c_1 |p_{HG}\rangle+c_2 |p_{BW}\rangle+c_3|p_{HW}\rangle+c_4|p_{BG}\rangle , \label{notproduct}
\end{equation}
where the unit vectors $|p_{HG}\rangle$, $|p_{BW}\rangle$, $|p_{HW}\rangle$ and $|p_{BG}\rangle$ represent the states $p_{\mathit{The\, Horse\, Growls}}$, $p_{\mathit{The\, Bear\, Whinnies}}$, $p_{\mathit{The\, Horse\, Whinnies}}$ and $p_{\mathit{The\, Bear\, Growls}}$, respectively. Eq. (\ref{notproduct}) is not, in general, a product, hence it is not equal to the tensor product $|p_{\mathit{Animal}}\rangle \otimes |p_{\mathit{Acts}}\rangle$, which is the mathematical basis of the presence of entanglement.

The unavoidability of entanglement could explain the difficulties that scholars encounter in putting forward a modeling scheme for concepts and their combinations in which individual concepts are represented by a unique mathematical structure, e.g., vectors such as in LSA, without introducing the tensor product structure (see, e.g., \cite{aertsczachor2004}).

\medskip
\noindent
{\bf Acknowledgments.} The authors are greatly indebted with the 81 friends and colleagues for participating in the experiment. This research was supported by Grants G.0405.08 and G.0234.08 of the Flemish Fund for Scientific Research.

\end{document}